\documentclass[authoryear,12pt]{elsarticle}

\usepackage{amssymb}
\usepackage{amsmath}

\usepackage{hyperref} 
\usepackage{subfigure} 
\usepackage{natbib}

\journal{arXiv}

\begin{document}

\begin{frontmatter}



\title{Simulation and Harmonic Analysis of \textit{k}-Space Readout (SHAKER)}


\author[a]{John C. Bodenschatz} 
\author[a,b]{Daniel. B. Rowe}

\affiliation[a]{organization={Department of Mathematical and Statistical Sciences, Marquette University},
            addressline={1250 W Wisconsin Ave}, 
            city={Milwaukee},
            postcode={53233}, 
            state={Wisconsin},
            country={USA}}
\affiliation[b]{country={Corresponding author email: daniel.rowe@marquette.edu}}

\begin{abstract}
In the realm of neuroimaging research, the demand for efficient and accurate simulation tools for functional magnetic resonance imaging (fMRI) data is ever increasing. We present SHAKER, a comprehensive MATLAB package for simulating complex-valued fMRI time series data that will advance understanding and implementation of the MR signal equation and related physics principles to fMRI simulation. The core objective of the package is to provide researchers with a user-friendly MATLAB graphical user interface (GUI) tool capable of generating complex-valued fMRI time series data. This tool will allow researchers to input various parameters related to the MRI scan and receive simulated \textit{k}-space data with ease, facilitating a deeper understanding of the intricacies of the generation and interpretation of fMRI data.
\end{abstract}




\end{frontmatter}



\section[Introduction]{Introduction} \label{sec:introduction}
\begin{figure}[b!]
    \centering
    \begin{subfigure}a.
        \centering
        \includegraphics[width=0.3\linewidth]{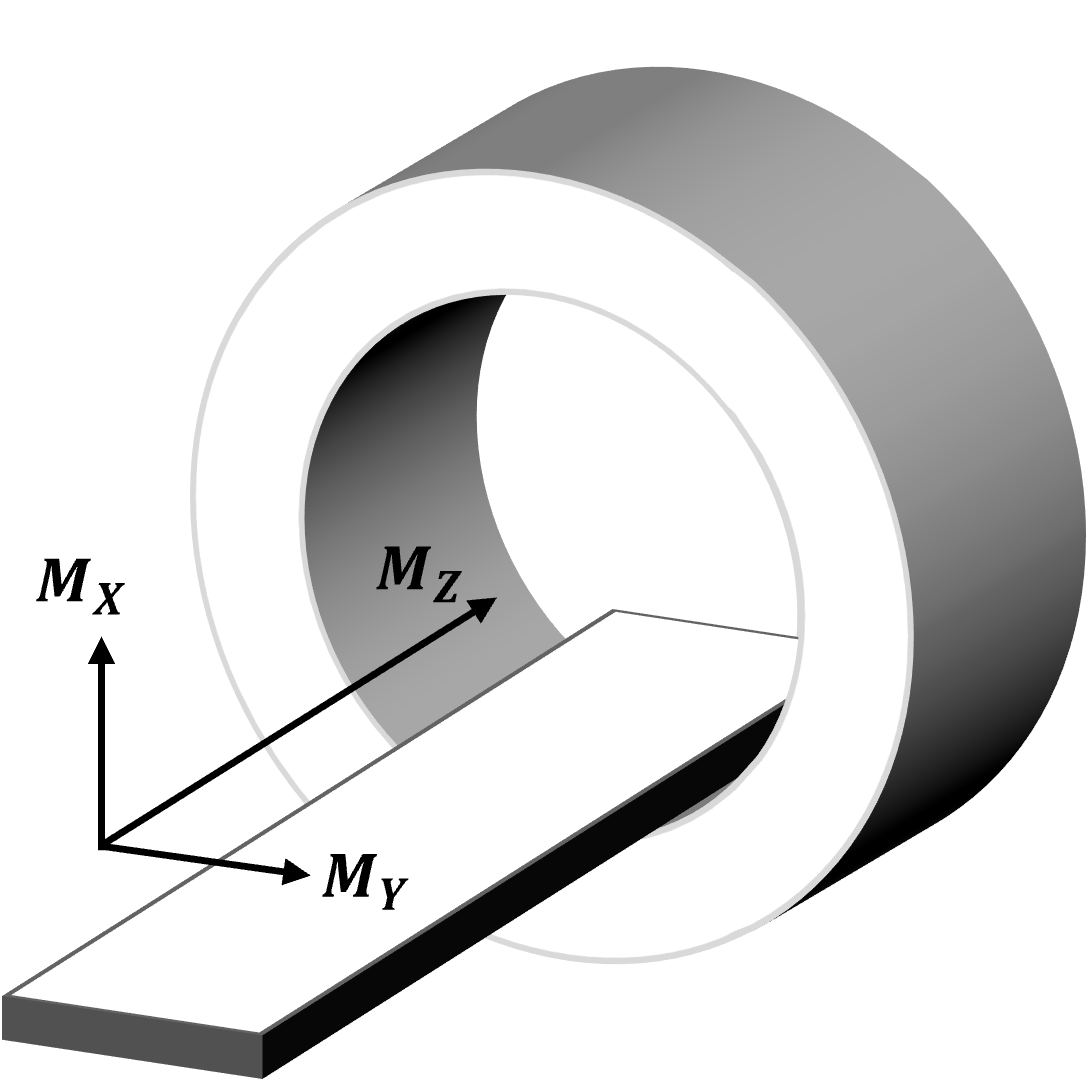}
        \label{fig:mri-machine}
    \end{subfigure}
    \begin{subfigure}b.
        \centering
        \includegraphics[width=0.25\linewidth]{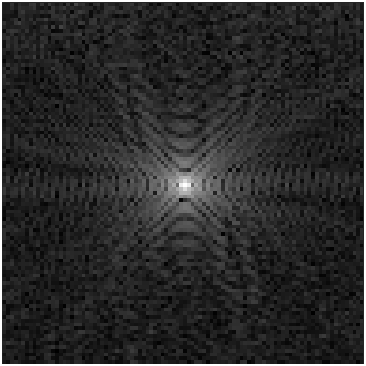}
        \label{fig:mri-kspace}
    \end{subfigure}
    \begin{subfigure}c.
        \centering
        \includegraphics[width=0.25\linewidth]{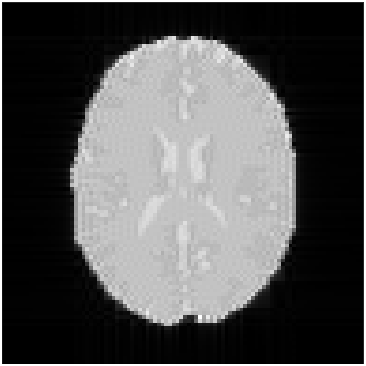}
        \label{fig:mri-image}
    \end{subfigure}
    \caption{The process of obtaining an image from the machine. a) MRI machine with main axes indicated; the $z$-direction is referred to as longitudinal, the $xy$-plane is referred to as the transverse plane. b) Acquired $k$-space array of spatial frequencies. c) Inverse discrete Fourier transform of $k$-space; a reconstructed image.} 
    \label{fig:mri-process}
\end{figure}

Functional magnetic resonance imaging (fMRI) is a non-invasive imaging technique that allows trained physicians and scientists to observe functionality of organs, in particular- the human brain. This is done by exciting protons in the various molecules that make up the different tissues of the organ, then determining a net change in magnetization as determined by an induced current in a loop of wire surrounding the patient. This net magnetization in different voxels of the region of interest (ROI) is associated with complex-valued spatial frequencies that fill $k$-space; a high order approximation of the Fourier transform of the voxel image of the organ. The $k$-space is then inverse discrete Fourier transformed (IDFT) to reconstruct an image. Figure \ref{fig:mri-process}a shows a simple MRI machine and the major axes, Figure \ref{fig:mri-process}b depicts an example of the magnitude of a measured complex-valued $k$-space array, and Figure \ref{fig:mri-process}c shows the magnitude of the complex-valued image reconstructed from complex-valued $k$-space, the magnitude of which is in Figure \ref{fig:mri-process}b.

To perform experiments in the machine is both financially and temporally costly; demanding machine time and obtaining Institutional Review Board (IRB) approvals can slow down the process of investigating new statistical techniques to extract information from fMRI data. Consequently, researchers will test developing methods on simulated data as a cost-effective way of measuring potential. Currently, simulated fMRI data are largely developed in-house for each researcher using a variety of methods. There has been work to develop a more standardized method to simulate fMRI time series data using various languages such as Python and R. However, many such methods disregard the complex-valued nature along with the true statistical and physical properties of the data output by the machine, in addition to returning magnitude-only images from simulations \citep{neurosim}. Some of these methods may also demand some form of in-line coding or require external files to support simulation \citep{snake-fmri}. It will be beneficial to provide a complete software tool to researchers that allows the simulation of complex-valued fMRI time series data with the ability to tune various parameters relating to the scan to match future experimental data, that will allow for proper testing of developing models. We present the current work on such an fMRI simulation software tool entitled \textit{Simulation and Harmonic Analysis of k-Space Readout} (SHAKER). SHAKER, a GUI-based simulator, is built on the physics-based principles of the MRI machine and is designed so that both new and well-versed researchers in the field can simulate data with ease. The sections that follow will give a brief overview of the physics being applied in the simulator, followed by an in-depth description of each of the parts of the simulator. This will be examined in an example simulation study at the end.

\subsection{Nuclear magnetic resonance} \label{subsec:nmr}
\begin{figure}[b!]
    \centering
    \begin{subfigure}a.
        \centering
        \includegraphics[width=0.25\linewidth]{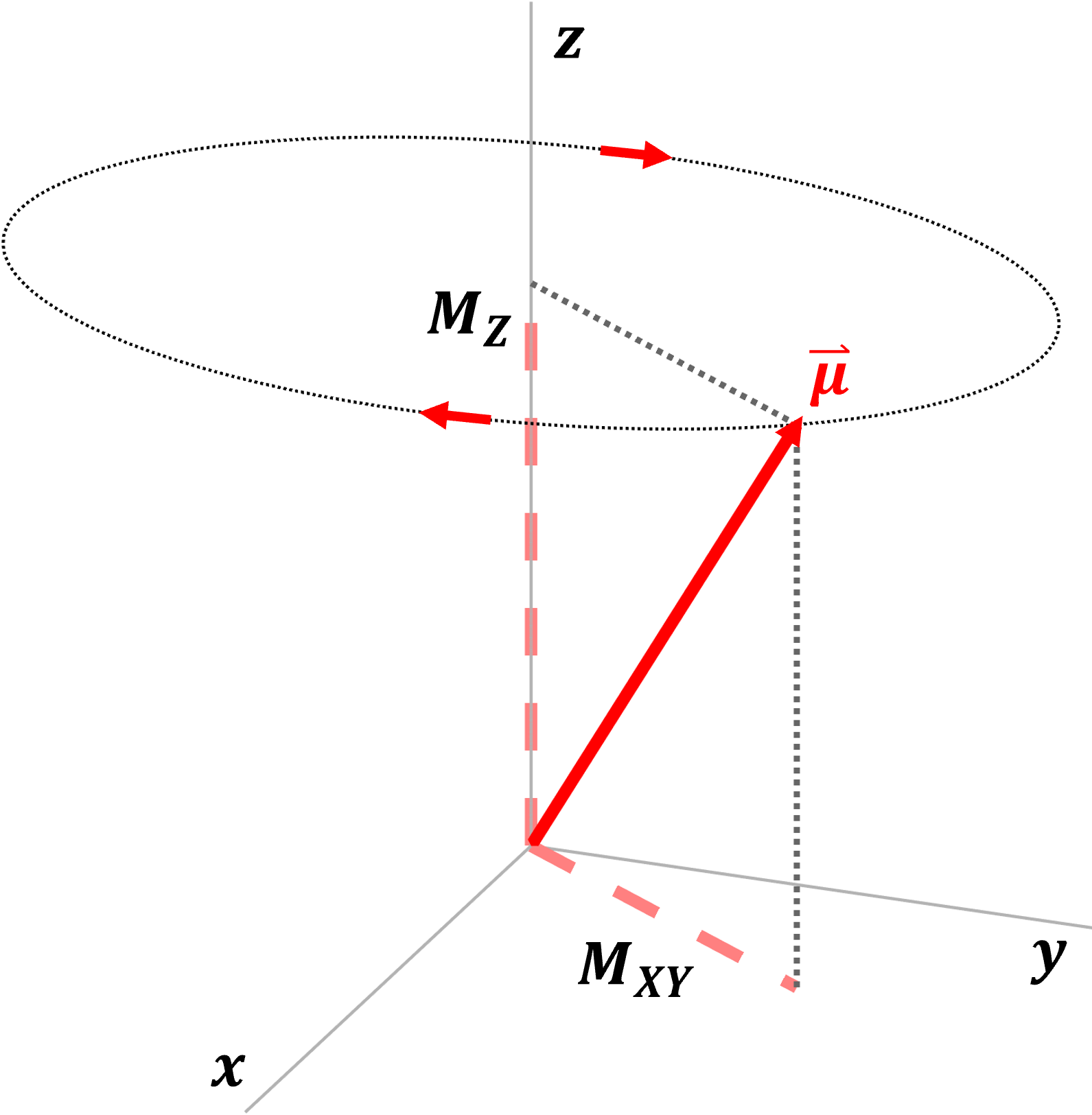}
        \label{fig:cone}
    \end{subfigure}
    \begin{subfigure}b.
        \centering
        \includegraphics[width=0.25\linewidth]{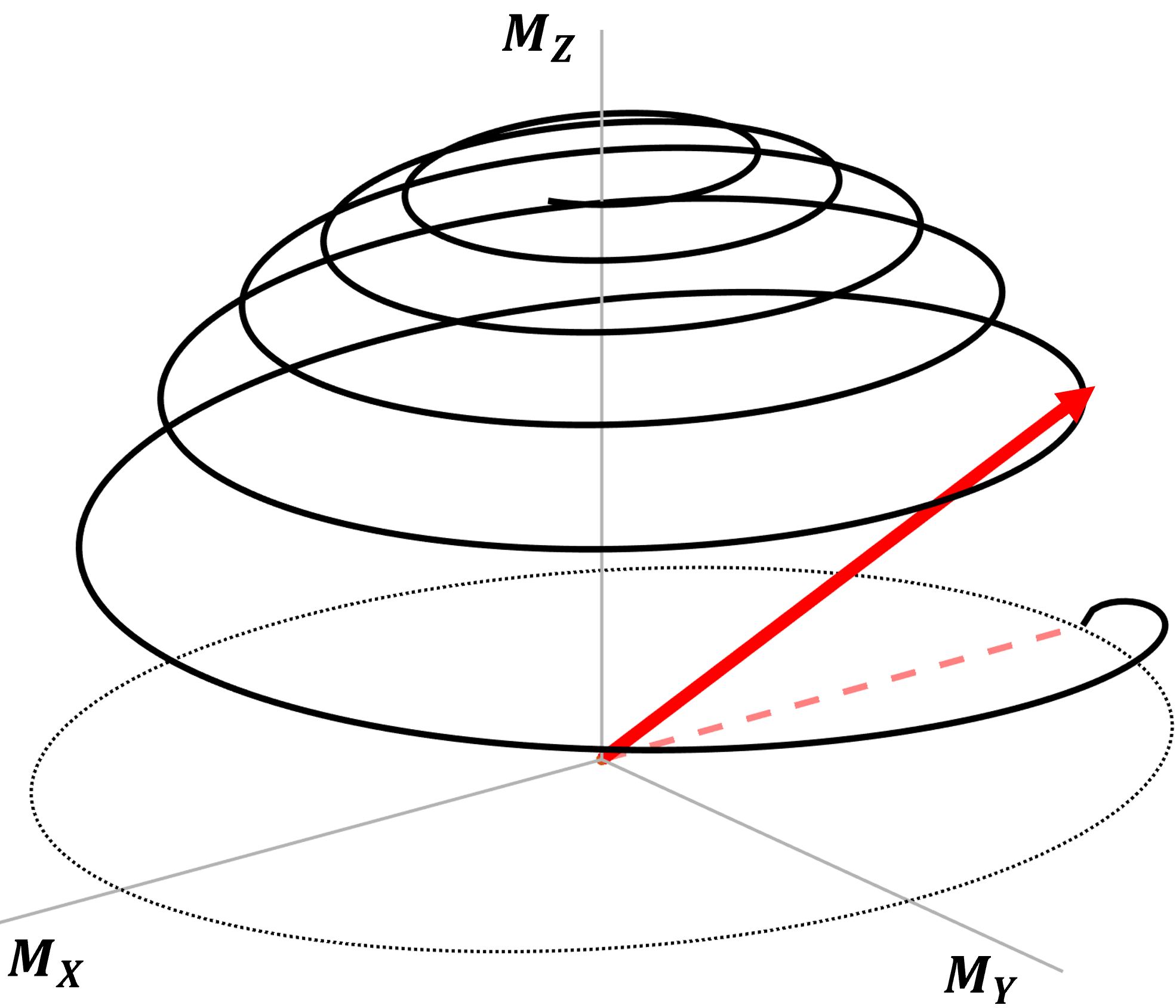}
        \label{fig:t1relaxation}
    \end{subfigure}
    \begin{subfigure}c.
        \centering
        \includegraphics[width=0.25\linewidth]{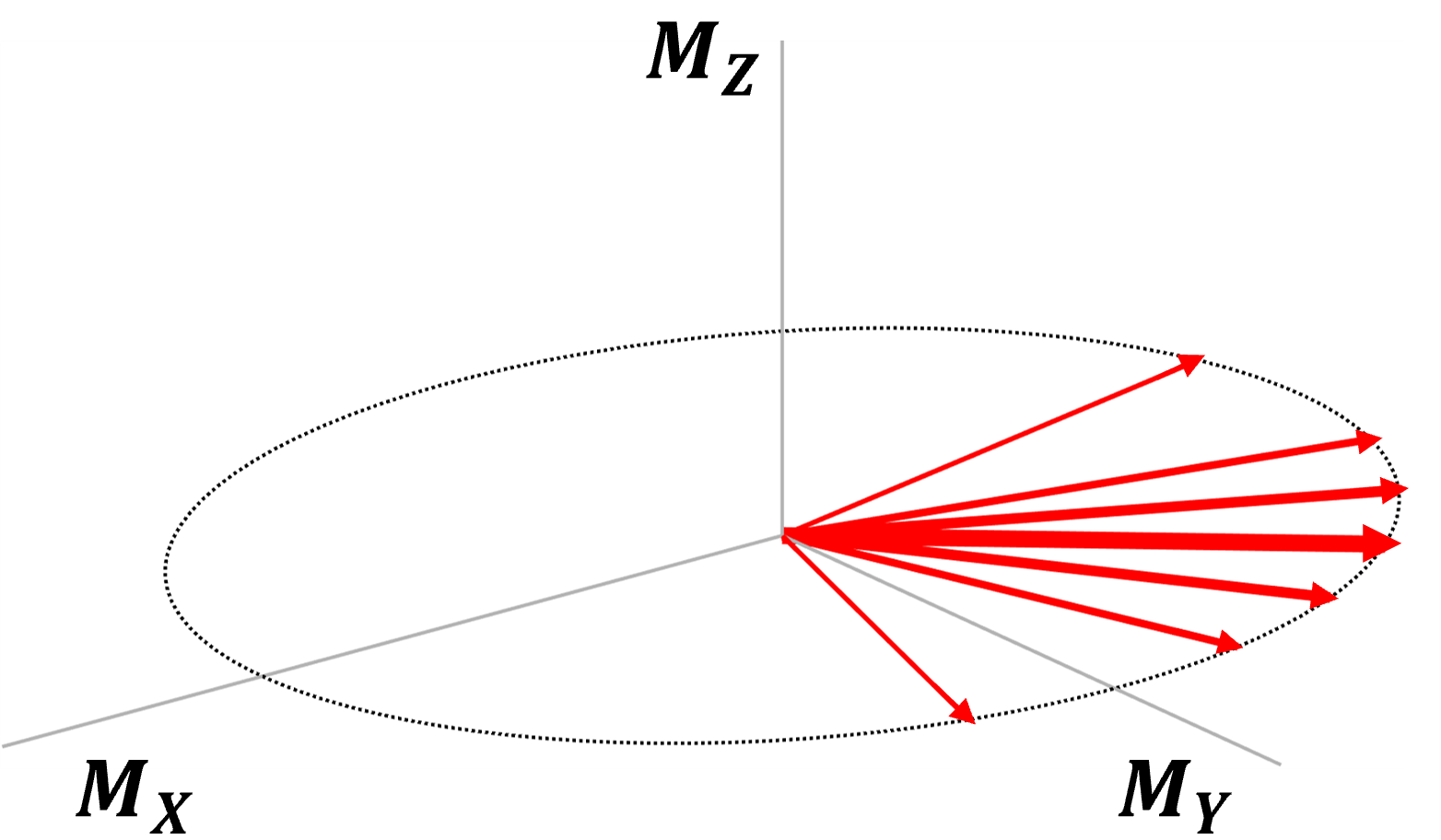}
        \label{fig:t2relaxation}
    \end{subfigure}
    \caption{Depicted in (a) is the net magnetization and precession of the magnetic moment about the central axis. Representations of the $T_1$ (b) and $T_2^{\ast}$ (c) relaxivities. $T_1$ is relaxation back into the longitudinal direction, $T_2$ is relaxation in the transverse direction.} 
    \label{fig:t1t2relaxation}
\end{figure}

A primary aim of SHAKER is to provide a \textit{realistic} simulation of fMRI. It is important to understand the physical principles and phenomena that determine the measured signal which is later reconstructed into an image. The MR machine creates a very strong magnetic field $B_0$ along the direction of the scanner as indicated in Figure \ref{fig:mri-process}a (1.5, 3, and 7 Tesla are common). This field aligns the spins of hydrogen nuclei within the body to become parallel with the direction of the scanner. The alignment of the hydrogen nuclei results in a net magnetization, denoted $M_0$. These hydrogen nuclei precess (resonate) at the Larmor frequency which is proportional to the external magnetic field they are exposed to,
\begin{equation}
    f_0 = \gamma B_0,
    \label{eqn:larmor}
\end{equation}
where $\gamma$ is the gyromagnetic ratio, a constant unique to each nucleus \citep{larmor}. In the case of hydrogen, we have $\gamma = 42.58MHz / T$. To excite these nuclei, a radio frequency (RF) burst of energy is sent into the system at this resonant frequency. The nuclei enter a higher energy state where their spins tip against the main magnetic field $B_0$ at some flip angle $\alpha$ determined by the length of the RF pulse. An $\alpha=90^\circ$ flip angle is common for fMRI. In the time that follows the RF pulse, these nuclei emit energy through two relaxation processes- $T_1$ and $T_2$. The longitudinal or spin-lattice relaxation time, $T_1$, is the recovery time for the parallel component of $M_0$, $M_Z$, back to equilibrium. The transverse or spin-spin relaxation time, $T_2$, is the decay of $M_{XY}$, the transverse component of $M_0$. In practice, $T_2^\ast$ is what is actually measured. The relationship between $T_2$ and $T_2^\ast$ is defined by
\begin{equation}
    1/T_2^\ast = 1/T_2 + 1/T_2^\prime,
    \label{eqn:t2star}
\end{equation}
where $1/T_2^\prime = \gamma \Delta B$ is the dephasing of the hydrogen nuclei as a result from hydrogen nuclei precessing at slightly different frequencies due to inhomogeneities in the magnetic field, $\Delta B$. The two effects, $T_1$ and $T_2^\ast$ are visualized in Figure \ref{fig:t1t2relaxation}b-c. Figure \ref{fig:t1t2relaxation}a shows the net magnetization change, a vector sum of the $T_1$ and $T_2^\ast$ relaxivities. These relaxivities result in a changing magnetic field within the tissue that is measured through current via Faraday's law of induction in one or more coils of wire that surround the bore of the machine. This measured signal is then later transformed into complex-valued images via the inverse discrete Fourier transform.

In fMRI, the blood-oxygen-level-dependent (BOLD) signal is interrogated to determine regions of activation \citep{BOLD}. The BOLD signal is a measure of localized brain blood level and oxygenation changes which are correlates for neuronal activity. These changes occur as a result of certain stimuli or tasks, e.g., right-hand finger tapping, that activate known regions of the brain. The BOLD signal presents itself as a $T_2^\ast$ effect since the change in magnetic properties of oxygenated and deoxygenated hemoglobin in blood causes a perturbance in the local magnetic field, $\Delta B$. Hence, fMRI time series are $T_2^\ast$-weighted images, highlighting regions of the brain with significant $T_2^\ast$ effects. 

Every image from an MRI machine comes from a predetermined ``pulse sequence'' of RF bursts and changing of magnetic gradients within the machine. In fMRI, images are most often collected via single shot echo planar imaging (EPI); ``single shot'' meaning only one RF excitation is applied per $k$-space array. The most commonly used pulse sequence used in EPI is gradient echo (GRE) \citep{KUMAR1975, handbook-mri-pulse-sequences}. In general, a given signal equation gives complex signal $s$ received at a given point $(k_x, k_y)$ in \textit{k}-space. The GRE signal equation is given by
\begin{equation}
        s(k_x, k_y) = \int\displaylimits_{-\infty}^\infty \int\displaylimits_{-\infty}^\infty \frac{M_0 \sin(\alpha)}{\left( 1-\cos(\alpha) e^{-TR/T_1} \right)} \left( 1-e^{-\text{TR}/T_1} \right) e^{-t/T_2^\ast}e^{i \gamma \Delta B t} e^{-i 2 \pi(k_x x + k_y y)} \,dx\,dy ,
    \label{eqn:GRE-signal}
\end{equation}
where $M_0(x,y)$, $T_1(x,y)$, $T_2^{\ast}(x,y)$, and $\Delta B(x,y)$ are functions of voxels $(x,y)$ within the physical object (or phantom) and $t(k_x,k_y)$ is the time at which the point $(k_x, k_y)$ in $k$-space is scanned. The simplification of replacing $t(k_x,k_y)$ with $TE$, echo time, is often used and is equivalent to assuming that all data are acquire at the $TE$ (SHAKER does not require this assumption). The repetition time, $TR$, is the time between successive RF pulses of the same slice, or equivalently, the time between successive measured $k$-space arrays of the same slice. The flip angle $\alpha$ is commonly set to $90^{\circ}$, which simplifies the first term in Equation \ref{eqn:GRE-signal} to just $M_0$. While GRE is most commonly used because of the high signal it provides, some higher strength scanners (7 T+) may opt to use the spin echo (SE) pulse sequence to detect BOLD signal \citep{Chen2015,nencka-rowe-SE-2005}. The SE signal equation is given by
\begin{equation}
    s(k_x, k_y) = \int\displaylimits_{-\infty}^\infty \int\displaylimits_{-\infty}^\infty M_0 \left( 1-e^{-\text{TR}/T_1} \right) e^{-t/T_2} e^{i \gamma \Delta B t} e^{-i 2 \pi(k_x x + k_y y)} \,dx\,dy ,
    \label{eqn:SE-signal}
\end{equation}
noting the use of $T_2$ instead of $T_2^\ast$. It has been shown that SE pulse sequences correct for the large scale dephasing caused by larger veins, which may not be as closely related to activation as capillaries \citep{kennan-spin-echo}. Closely related to the SE pulse sequence, but not generally used for fMRI experiments, is the inversion recovery (IR) pulse sequence. The signal equation for IR is given by 
\begin{equation}
    s(k_x, k_y) = \int\displaylimits_{-\infty}^\infty \int\displaylimits_{-\infty}^\infty M_0 \left( 1-2e^{-\text{TI}/T_1} + e^{-\text{TR}/T_1} \right)  e^{i \gamma \Delta B t} e^{-i 2 \pi(k_x x + k_y y)} \,dx\,dy ,
    \label{eqn:IR-signal}
\end{equation}
where TI is the inversion time. The IR pulse sequence is more commonly used for $T_1$-weighted images, as compared to the $T_2^\ast$-weighted image that is standard in fMRI, but is still included in SHAKER. Both the SE and IR signal equations assume a $90^\circ$ flip angle from the initial RF pulse. Other pulse sequences such as diffusion weighted imaging (DWI-fMRI), and saturation recovery (SR) may be included in the future plans for SHAKER development.

\subsection{$k$-Space and the Fourier transform} \label{subsec:kspace}
\begin{figure}[b!]
\centering
\includegraphics[width=1\linewidth]{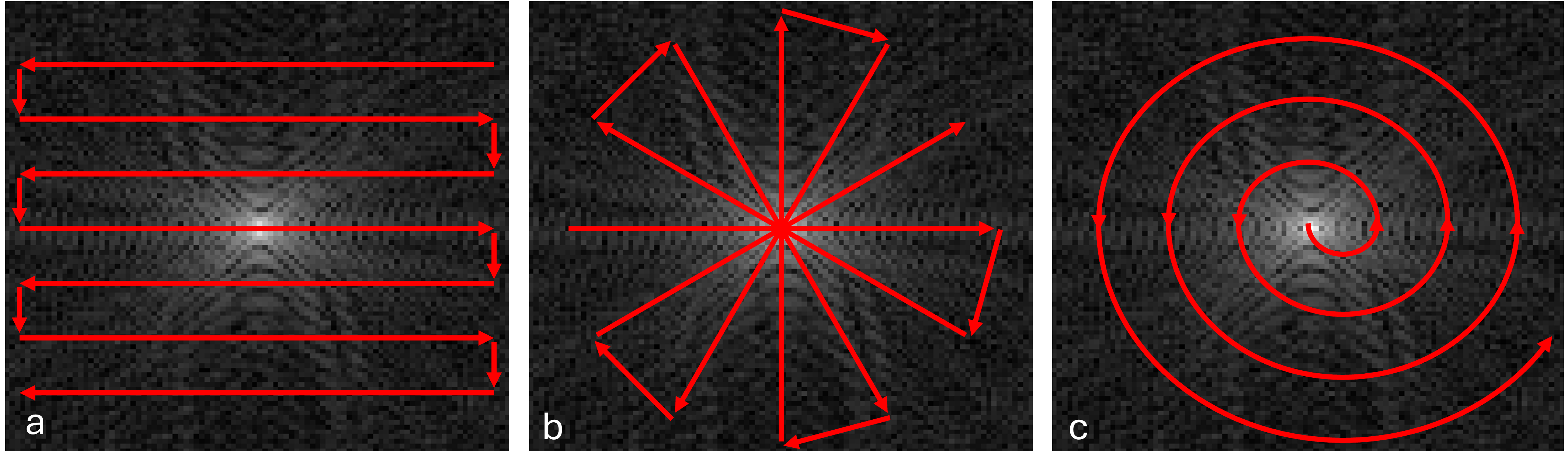}
\caption{\label{fig:trajectories}Three examples of $k$-space trajectory. a) Standard Cartesian encoding. b) Radial encoding. c) Spiral encoding.}
\end{figure}

The signal equations from Section \ref{subsec:nmr} are measured in the spatial-frequency domain called $k$-space (magnitude images of $k$-space are presented in Figure \ref{fig:trajectories}). Each of Equations \ref{eqn:GRE-signal}-\ref{eqn:IR-signal} could be condensed to
\begin{equation}
    s(k_x, k_y) = \int\displaylimits_{-\infty}^\infty \int\displaylimits_{-\infty}^\infty \rho_0 (x,y)  e^{-i 2 \pi(k_x x + k_y y)} \,dx\,dy .
    \label{eqn:general-signal-continuous}
\end{equation}
In this form, we can see that the signal equation is the Fourier transform of $\rho_0$, the net magnetization after having been weighted by the relevant relaxivities. In practice, however, $k$-space is only measured at a finite set of discrete points. So, we can discretize Equation \ref{eqn:general-signal-continuous} into
\begin{equation}
    s(k_x, k_y) = \frac{1}{N_x N_y}\sum\displaylimits_{m=0}^{N_{x}-1} \sum\displaylimits_{n=0}^{N_{y}-1} \rho_0 (x,y)  e^{-i 2 \pi \left(\frac{k_x}{N_x} x_m + \frac{k_y}{N_y} y_n \right)} ,
    \label{eqn:dft}
\end{equation}
where $N_x$ and $N_y$ are the number of points in image-space in the $x$ and $y$ directions, respectively. In fMRI, it is common that $N_x = N_y = 64,\;96,\;128$. Thus, we arrive at $k$-space equating to the \textit{discrete} Fourier transform of image-space.

The objects and phantoms being imaged are composed of real-valued voxels only. So, ideally, the Fourier transform of these objects would result in a $k$-space that maintains Hermitian symmetry. However due to the terms involving $T_2^{\ast}$ and $\Delta B$, the expected Hermitian symmetry of $k$-space is broken. In fact, when looking at Equations \ref{eqn:GRE-signal}-\ref{eqn:IR-signal}, the inclusion of the these terms necessarily implies that $k$-space is only a very close \textit{approximation} to the Fourier transform of image-space, since the terms are time-dependent. This leads to possible distortions and artifacts when reconstructing $k$-space into images using the inverse discrete Fourier transform
\begin{equation}
    \rho_0 (x,y) = \frac{1}{N_{k_x} N_{k_y}}\sum\displaylimits_{m=0}^{N_{k_x}-1} \sum\displaylimits_{n=0}^{N_{k_y}-1} s(k_x, k_y)  e^{-i 2 \pi \left(\frac{x}{N_{k_x}} {k_x}_m + \frac{y}{N_{k_y}} {k_y}_n \right)} ,
    \label{eqn:idft}
\end{equation}
however it is still the most common method of image reconstruction in MRI. The incorporation of prior knowledge regarding the relaxivities as well as magnetic field inhomogeneity has been implemented to enhance image reconstruction \citep{KaramanBruceRoweMRI15}.

As a result of physical limitations, $k$-space must be scanned, or traversed, in one continuous path. The most conventional method is to scan horizontal rows, often referred to as the frequency-encoding direction, in alternating directions working up (or down) $k$-space in the phase-encoding direction as shown in Figure \ref{fig:trajectories}a. This involves a set of ``turnaround points'' at the end of each row that are often discarded or not measured, resulting in dead scan time that decreases the rate of useful data acquisition. This has proven to be a convenient way to scan $k$-space as it results in a Cartesian encoding of the spatial frequencies which allows for the simple inverse discrete Fourier transform to reconstruct $k$-space back into an image. Other $k$-space trajectories, including non-Cartesian methods such as PROPELLER, radial as in Figure \ref{fig:trajectories}b (equivalent to PROPELLER with blade width 1), and spiral as in Figure \ref{fig:trajectories}c have been implemented for various reasons such as reducing scan time and increasing robustness to artifacts due to motion \citep{PROPELLER,spiral}. 




\section{Design} \label{sec:design}

\begin{figure}[b!]
\centering
\includegraphics[width=0.75\linewidth]{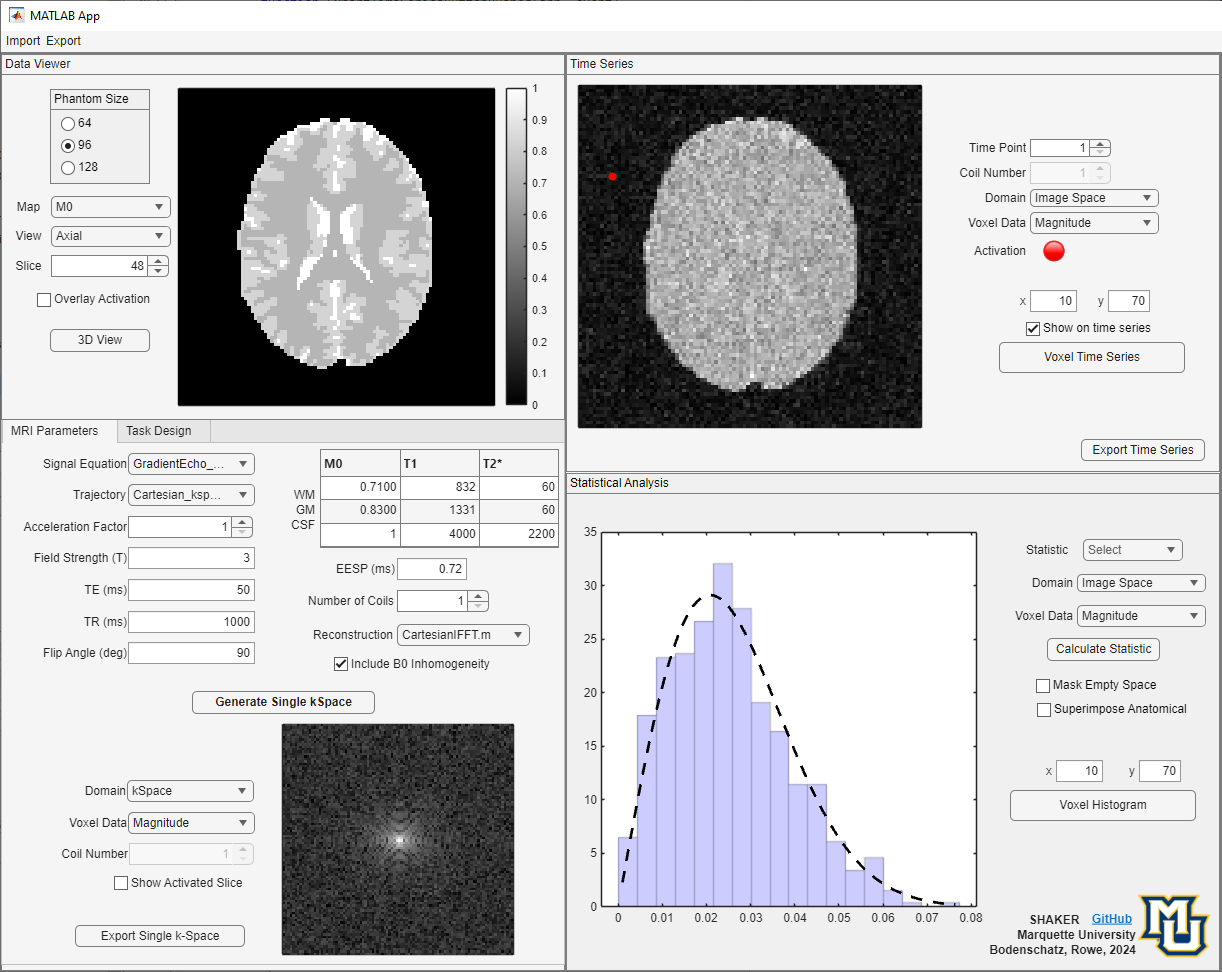}
\caption{\label{fig:SHAKER_GUI} A screenshot of the working SHAKER GUI. In the top left pane, users can view the raw phantom data that will be input to the signal equation. In the bottom left pane are the adjustable MRI parameters and the fMRI experimental setup. The right pane displays two views: the top presents unaltered data from the simulated time series, and the bottom reflects an example of a statistical data set created from the time series data (in this case: a histogram of one voxel's magnitude time series).}
\end{figure}

SHAKER is an all-inclusive fMRI simulation software package built with the user in mind. SHAKER is built using the MATLAB programming language and presents as a GUI (Figure \ref{fig:SHAKER_GUI}), with no scripting or external data required \citep{MATLAB}. In the top left pane of Figure \ref{fig:SHAKER_GUI}, users can view the pre-loaded digital phantom. The bottom left pane involved customization of MRI parameters and fMRI experimental design options. The top right pane presents a view of the simulated time series data. The bottom right pane is where statistical maps and measurements from the time series data can be observed. This is also where any models in development may be tested on the simulated data. All code and data used to operate SHAKER are publicly available on GitHub to encourage a better understanding and allow customizations to be made. The contents of this section explain in detail the functionality of SHAKER in each of these panes.

\subsection{Digital phantom} \label{subsec:data}
SHAKER comes pre-loaded with a full volume digital phantom that was simulated with realistic $M_0$, $T_1$, and $T_2^\ast$ values based on a 3 T machine \citep{Karaman_2014_phd}. The tissues included in the phantom are gray matter (GM), white matter (WM), and cerebrospinal fluid (CSF). The $\Delta B$ map was considered as a gradient along each of the dimensions of the scanner, combined with some biological detail from the $T_2^\ast$ map. This included digital phantom is stored as a MATLAB structure array: \texttt{Phantom: M0, T1, T2, deltaB}, and users may also load in their own maps.

\begin{figure}[b!]
\centering
\includegraphics[width=1\linewidth]{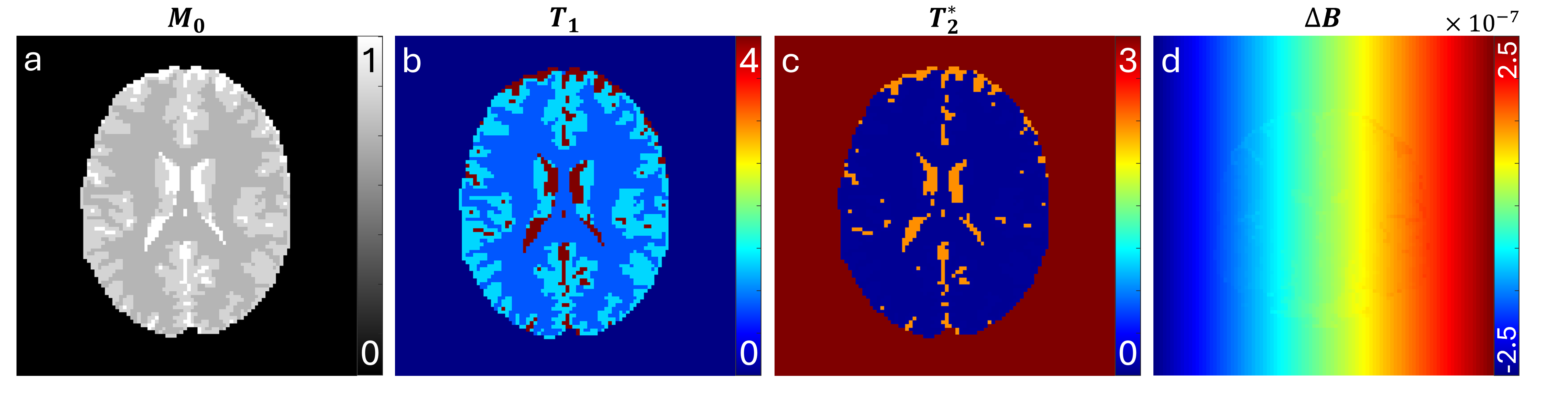}
\caption{\label{fig:maps} Maps taken from an axial slice of the digital phantom. a) Net Magnetization, $M_0$ (dimensionless). b) Longitudinal relaxation, $T_1$ (seconds). c) Transverse relaxation, $T_2^\ast$ (seconds). d) Field inhomogeneity, $\Delta B$ (Tesla).}
\end{figure}

An axial slice of the phantom is shown in Figure \ref{fig:maps}. In some cases, a higher or lower sampling density requirement is needed in $k$-space. To support this, the digital phantom can be rescaled to $64\times64\times64$ or $128\times128\times128$ by changing the \textit{Phantom Size} option. Should one want to implement their own digital phantom, it will be necessary to create a structure with the same naming convention, having four maps whose dimensions all agree. This custom phantom can then be imported as a \texttt{.mat} file from the toolbar located at the top of the GUI. Additionally, there is an activation map included with SHAKER that carries the same dimension as the phantom. The activation map is a binary array, with ones only at the intended location(s) of simulated activation. It is designed such that it roughly resembles the left primary motor cortex region of the brain- the area that is expected to be active during right-hand finger tapping. Custom activation maps can also be imported as \texttt{.mat} files with type \texttt{double} and name \texttt{ActMap} and should share dimensions with the phantom being used for simulation.

\begin{figure}[h!]
\centering
\includegraphics[width=0.5\linewidth]{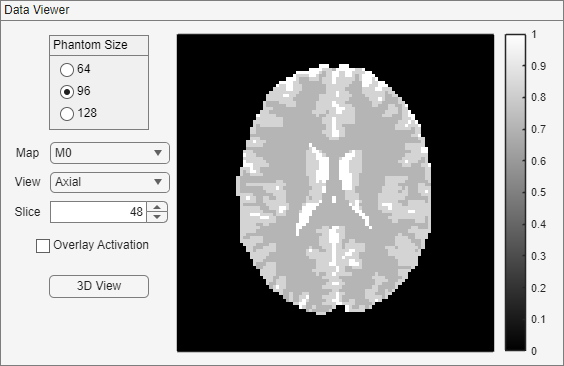}
\caption{\label{fig:data-example} \textit{Data Viewer}: the top left pane of SHAKER. Here the phantom size can be selected, each of the four maps can be viewed, and slice orientation/location can be selected.}
\end{figure}

At present, SHAKER is equipped to handle single-slice excitations in any of the three major planes: axial, sagittal, or coronal. Support for echo-volume imaging could be supported in future versions. Slice selection and orientation are both chosen and viewed in the top left pane of the GUI, titled \textit{Data Viewer} as in Figure \ref{fig:data-example}. The size of the phantom is also adjusted from this pane. Two other options that have no effect on the simulation: viewing each of the maps and visualizing where the activation is expected, are available from this panel as well. Making use of the recently developed \texttt{volshow()} function in MATLAB, users can get a 3-D view of the four maps that make up the phantom, sliced at the indicated location and orientation.

\subsection{MRI parameters} \label{subsec:mri}
\begin{figure}[b!]
\centering
\includegraphics[width=0.5\linewidth]{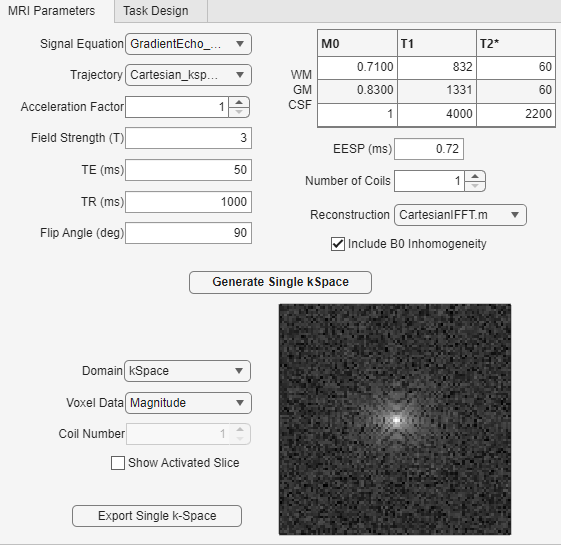}
\caption{\label{fig:mri-example} \textit{MRI Parameters}: the first tab of the bottom left pane of SHAKER. Here relevant MRI parameters can be set. A single $k$-space can be simulated and observed to check that settings are correct before simulating an entire time series.}
\end{figure}

Found in the bottom left pane of the GUI are tabs for \textit{MRI Parameters} and \textit{Task Design} as shown in Figures \ref{fig:mri-example} and \ref{fig:fmri-example}. \textit{MRI Parameters} is the part of SHAKER where users will make selections similar to that of an MRI technician. \textit{Task Design} will be discussed in Section \ref{subsec:fmri}. All settings in this pane should be set before initializing any simulations. From the top toolbar of SHAKER, the MRI structure being used for simulation can be saved into a \texttt{.mat} file which contains a MATLAB structure array named \texttt{MRI}. This file may be imported to future instances of SHAKER for ease of reproducibility of results.

The first two options for MRI parameters are \textit{Signal Equations} and \textit{Trajectory}. These refer to a choice of signal equation as described in Section \ref{subsec:nmr} and a $k$-space trajectory from Section \ref{subsec:kspace}. The $k$-space trajectory functions are designed to receive the \texttt{MRI} object inherent to SHAKER that contains all pertinent information regarding the scanner properties. The trajectory functions then return three arrays: one each for the $k_x$ and $k_y$ locations at which $k$-space is sampled as well as an array noting the time at which the points are sampled $t(k_x, k_y)$. These arrays are stored in an object within SHAKER for later reference. Following this, the user can select their choice of signal equation function, which receives input data about the phantom, $k$-space sampling, and the \texttt{MRI} object, then return a simulated array of $k$-space measurements. Both the signal equation and $k$-space trajectory are two files that users can create their own version of, using the templates provided by SHAKER, to sample $k$-space in their own preferred way. Further details regarding inputs/outputs can be found in the appendix.

The next option is \textit{Acceleration Factor}. This can mean different things depending on the context of the $k$-space sampling method. For example, in Cartesian trajectories of $k$-space an acceleration factor of $n_a$ is often implemented as a measurement of every $n_a$ lines in the frequency-encode direction. In the single-spoke radial trajectory of $k$-space, this is commonly the measurement of every $n_a$ spokes. There is no restriction on how this might be implemented in one's own $k$-space trajectory file. Since SHAKER currently supports single-slice imaging, this acceleration factor should be interpreted as an \textit{in-plane acceleration} (IPA). Following this is the choice to change the simulated magnetic field strength. The values found in the table located at the top right of Figure \ref{fig:mri-example} can be altered to produce an effect on other $k$-space features at the will of the user. The four options that follow, \textit{TE}, \textit{TR}, \textit{Flip Angle}, and \textit{EESP} are direct inputs to the signal equation as described in Section \ref{subsec:nmr}.

\begin{figure}[t!]
\centering
\includegraphics[width=65mm,scale=0.5]{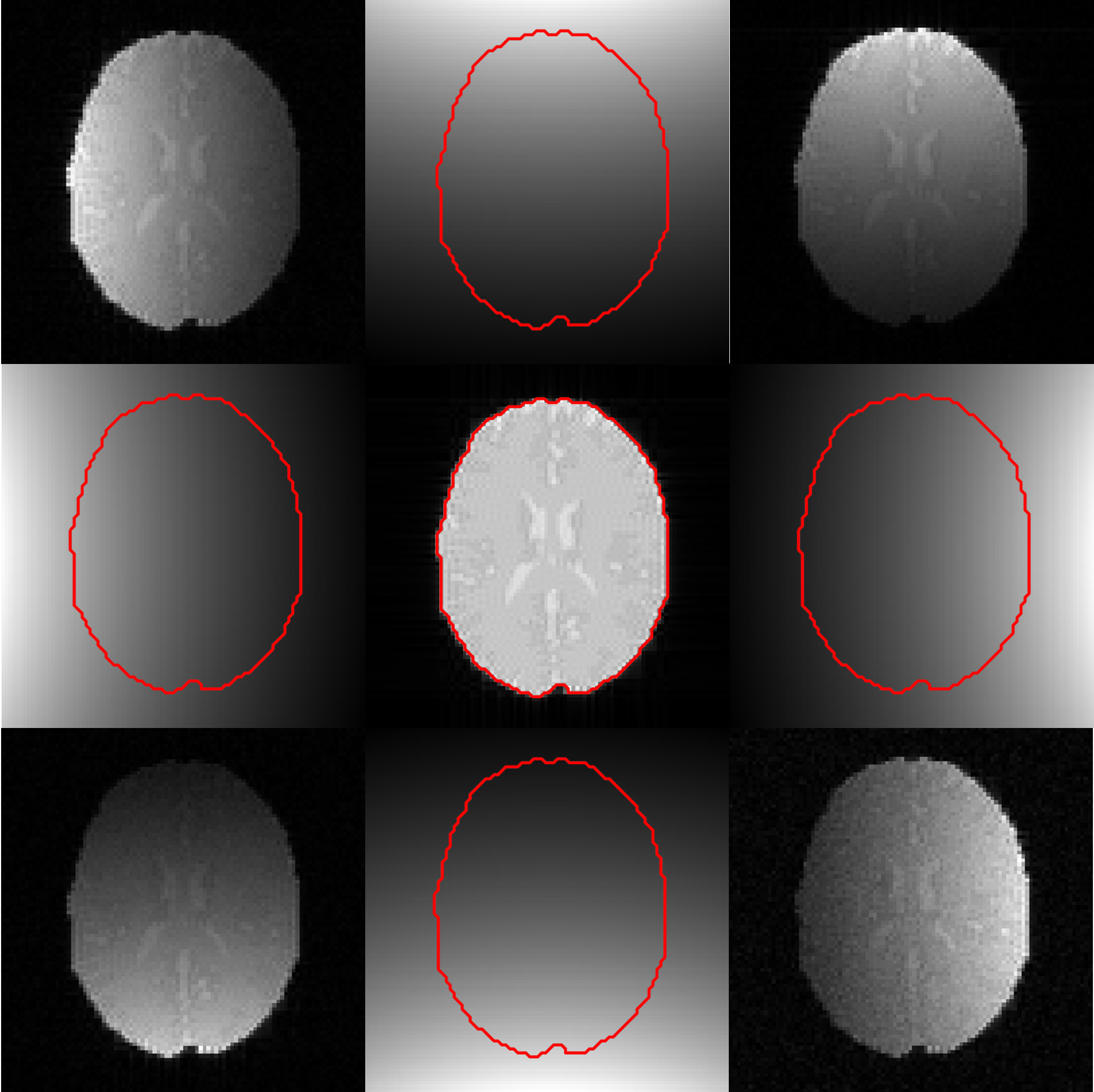}
\caption{\label{fig:coils} Coil sensitivities (top, right, bottom, left) and sensitivity weighted images (corners) for the case of $n_c=4$. The center image is the true, unweighted slice image. The red outlines indicate the location of the slice as seen by each of the coil sensitivities.}
\end{figure}

Parallel imaging in fMRI has received a lot of attention recently due to it's ability to accelerate the rate at which images are acquired in fMRI experiments \citep{sense,grappa}. In practice, each coil measures a sensitivity-weighted image of the brain, or phantom, at no additional temporal cost. SHAKER supports the use of a single, uniform coil, or multiple coils aligned with the bore of the machine. Users may specify any number of coils to simulate their data by changing the value for the \textit{Number of Coils} option. The $n_c$ coils have sensitivity matrices that match the dimensions of the phantom. In SHAKER, each of the $n_c$ coil sensitivities is constructed by placing a point at a each coil location, all of which are equidistant from the center of image-space and angularly equidistant from each other. The sensitivity of each coil array then decreases proportional to the inverse of the distance from this point. An example of the simulated coils, coil sensitivity weighted images, and averaged image is shown in Figure \ref{fig:coils}.

The last option is the choice of \textit{Reconstruction Algorithm}. While SHAKER is a $k$-space simulation tool, the reconstruction of images is supported for the more common $k$-space trajectories. Similar to the $k$-space trajectory and signal equation, the choice of reconstruction algorithm can be user-created based on an included template. These templates can be found in the subdirectories for each of the respective steps in the simulation process. More detail on this can be found in the Appendix.

\subsection{FMRI experimental design and noise}  \label{subsec:fmri}
\begin{figure}[t!]
\centering
\includegraphics[width=0.5\linewidth]{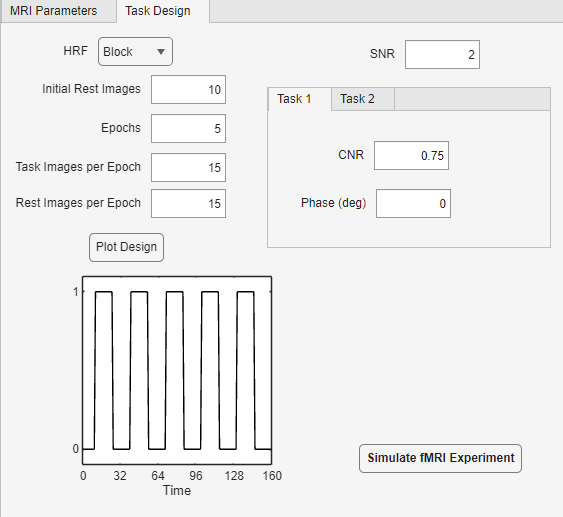}
\caption{\label{fig:fmri-example} \textit{Task Design}: the second tab of the bottom left pane of SHAKER. This is where users may adjust the experimental design of the time series, as well as specify \textit{SNR} and \textit{CNR}.}
\end{figure}

Task-based fMRI generally starts with an initial set of rest images that allow the tissue to reach a steady state in the magnetic field \citep{questions}. Following this, many epochs of rest / task images are taken. For example, an experiment may include: 16 initial rest images followed by 19 epochs of 16 task images followed by 16 rest images for a total of 624 images. It is often the case that some or all of the initial rest images are discarded for fMRI analysis due to the fact that they yield a higher signal than the steady state images. This can be circumvented by increasing the flip angle for the first few images so that the amount of transverse magnetization excited in each image is approximately the same \citep{FLASH}. For this example, discarding the first 16 from analysis would give $n_{IMG}=608$ images in the fMRI time series. It has been shown, however, that the first few images can be used to aid in analysis of the measured fMRI data, e.g., $T_1$ map estimation. In SHAKER, users may choose a set number of initial rest images, the number of epochs, and number of rest/task images per epoch. This is then stored as a design vector that can be used for later analysis of the simulated time series. This is all done from the second tab of the bottom left pane of SHAKER, as in Figure \ref{fig:fmri-example}.

Complex-valued voxel measurements $k_c$ in fMRI are composed of both a real and imaginary part, $k_c = k_R + ik_I$ where $i=\sqrt{-1}$. The measured magnitude $r_k$ and phase $\phi_k$ of the voxels come from the transformation $k_R = r_k \cos(\phi_k)$ and $k_I = r_k \sin(\phi_k)$. To better model the process of the machine, SHAKER adds noise to $k$-space directly rather than to reconstructed images as is often done. Since the analog-to-digital converters (ADCs) collect $k$-space measurements independently, it is understood that the real and imaginary parts of $k$-space measurements are independent and identically distributed (\textit{iid}) normally for each spatial frequency. Thus, the joint distribution is given as
\begin{equation}
f(k_R, k_I) = \frac{1}{(2\pi\sigma_{k}^2)^{1/2}}\exp\left[ -\frac{(k_R - \mu_{k,R})^2}{2\sigma_{k}^2} \right] \frac{1}{(2\pi\sigma_{k}^2)^{1/2}}\exp\left[ -\frac{(k_I - \mu_{k,I})^2}{2\sigma_{k}^2} \right],
\label{eqn:k-space-dist}
\end{equation}
where  $\mu_{k,R}$ and $\mu_{k,I}$ are the \textit{true} real and imaginary components of the spatial frequency \citep{henkelman,noise}. The inverse Fourier transform of the real and imaginary components of the noise from $k$-space into image space will also be normally distributed with a scaled variance. This relationship is given by
\begin{equation}
     \sigma^2 = \frac{\sigma_{k}^2}{n_x  n_y},
    \label{eqn:variance-scale}
\end{equation}
where $\sigma^2$ is the variance of the normally distributed real/imaginary noise in image space, $\sigma_{k}^2$ is the variance of the normally distributed real/imaginary noise in $k$-space, and $n_x$, $n_y$ are the dimensions of reconstructed image space \citep{handbook-chapter-rowe}. This fact reveals that the joint distribution of a voxels real and imaginary parts in image-space can be written similarly to Equation \ref{eqn:k-space-dist} as
\begin{equation}
f(y_R, y_I) = \frac{1}{(2\pi\sigma^2)^{1/2}}\exp\left[ -\frac{(y_R - \mu_{R})^2}{2\sigma^2} \right] \frac{1}{(2\pi\sigma^2)^{1/2}}\exp\left[ -\frac{(y_I - \mu_{I})^2}{2\sigma^2} \right].
\label{eqn:image-space-dist}
\end{equation}
The true real and imaginary components, $\mu_r$ and $\mu_I$, can be expressed in terms of the true magnitude and phase, $\rho$ and $\theta$, by the transformation $\mu_R = \rho\cos(\theta)$ and $\mu_I = \rho \sin(\theta)$. Since the magnitude of voxels is preferred over the real/imaginary values when looking at an image, we can transform the measured random variables $(y_R, y_I)$ to $(r, \phi)$ where $y_R = r \cos(\phi)$ and $y_I = r \sin(\phi)$ \citep{rowe-jsm-2023}. Calculating the Jacobian to be $J=r$, this gives the joint distribution
\begin{equation}
    f(r,\phi) = \frac{r}{2\pi\sigma^2} \exp \left(  -\frac{1}{2\sigma^2} \left[    r^2 + \rho^2 - 2r\rho \cos(\phi-\theta)     \right]     \right).
    \label{eqn:joint}
\end{equation}
By integrating out $\phi$ from Equation \ref{eqn:joint}, we get a Ricean marginal distribution for the voxel's magnitude $r$ \citep{rice-distribution,rician-noise,rowe-phase-2005,adrian-maitra-rowe-stat13},
\begin{equation}
    f(r) = \frac{r}{\sigma^2} \exp \left[ - \frac{r^2+\rho^2}{2\sigma^2} \right] I_0 \left( \frac{r\rho}{\sigma^2} \right).
    \label{eqn:rice}
\end{equation}
Here $I_0$ is the zeroth order modified Bessel function of the first kind. The mean of the Ricean distribution is $\sigma \sqrt{\pi/2} L_{1/2} (-\rho^2 / 2\sigma^2)$ where $L_{1/2}$ is a Laguerre polynomial. The variance of the Ricean distribution, denoted as $\sigma_r^2$, has the following relationship with the variance of the real and imaginary components of voxels in image space, $\sigma^2$,
\begin{equation}
    \sigma_r^2 = 2\sigma^2 + \rho^2 -\frac{\pi \sigma^2}{2} L_{1/2}^2 (-\rho^2 / 2\sigma^2).
    \label{eqn:rice-variance}
\end{equation}
The subscript $r$ is used to indicated the observed magnitude. In regions of empty space where the true signal $\rho$ is small, $\rho\approx 0$, this is reduced to the Rayleigh distribution with mean $\sigma \sqrt{\frac{\pi}{2}}$ and variance $\frac{4-\pi}{2}\sigma^2$ \citep{rayleigh-distribution}. In regions of space with high true signal $\rho$, this becomes the normal distribution with mean $\rho$ and variance $\sigma^2$. Integrating out the magnitude $r$ from the joint distribution in Equation \ref{eqn:joint} gives the unnamed non-normal distribution marginal distribution for the phase $\phi$,
\begin{equation}
    f(\phi) = \frac{1}{2\pi} \exp \left[ - \frac{\rho^2}{2\sigma^2}  \right] \left[  1 + \frac{\rho}{\sigma} \sqrt{2\pi}\cos(\phi - \theta) \exp \left[  \frac{\rho^2 \cos^2(\phi-\theta)}{2\sigma^2}  \right]  \Phi \left( \frac{\rho \cos(\phi - \theta)}{\sigma}  \right) \right],
    \label{eqn:non-normal}
\end{equation}
where $\Phi(x)$ is the cumulative distribution function of the standard normal distribution. When the signal $\rho$ is near zero, the phase will be uniformly distributed on $\left[-\pi, \pi\right]$ with mean $0$ and variance $\frac{\pi^2}{3}$. When the signal $\rho$ becomes large, the distribution of the phase becomes normal with mean $\theta$ and variance $\frac{\sigma^2}{\rho^2}$.

Task-based fMRI for an individual voxel's magnitude time series $r_t$ can be expressed as the linear equation
\begin{equation}
    r_t = \beta_0 + \beta_1 x_t + \varepsilon_r.
    \label{eqn:fmri-linear}
\end{equation}
As previously recognized, the additive noise is Ricean distributed with variance $\sigma_r^2$ from Equation \ref{eqn:rice-variance}. Here, $\beta_0 \in \mathbf{R}$ is the baseline signal which determines the signal-to-noise ratio $SNR = \beta_0 / \sigma_r$, and $\beta_1 \in \mathbf{R}$ is the task-related signal increase which determines the contrast-to-noise ratio $CNR = \beta_1 / \sigma_r$. The design vector $x_t \in \{0,1\}^{n_t}$ has length equal to the number of reconstructed images in the time series, $n_t$. In $x$, indices corresponding to a non-task image have an element of 0, while indices corresponding to a task-active image have an element of 1; this is equivalent to a block design hemodynamic response function (HRF). At present, SHAKER supports block design HRFs only. Future development considerations include the option to convolve user-defined functions with the block design to create custom HRFs.

When multiple coils are introduced to support parallel imaging, the statistical properties of the data are altered. Considering $C$ receiver coils, the composite magnitude signal is given by $M_C=\sqrt{\Sigma_{j=1}^{C} \left[   M_{jR}^2 + M_{jI}^2     \right]}$ where $M_{jR}=\rho_{jC}\cos(\theta_{jC})$ and $M_{jI}=\rho_{jC}\sin(\theta_{jC})$ denote the real and imaginary signals reconstructed from the $j^{th}$ receiver coil. It has been shown that the probability density of this composite magnitude is the non-central Chi distribution, which can be written as:
\begin{equation}
        f(M_C) = \frac{\rho_C}{\sigma^2}  \left( \frac{M_C}{\rho_C}  \right)^C    \exp\left(  -\frac{(\rho_C^2 + M_C^2)} {2\sigma_2}  \right) I_{C-1}\left(  \frac{M_c \rho_C}{\sigma^2} \right)
    \label{eqn:non-center-chi}
\end{equation}

\noindent where $\rho_C$ is the true signal magnitude \citep{Koay2006}.

When designing an fMRI experiment in SHAKER, users may specify an \textit{SNR} and \textit{CNR} which will be used to calculate the baseline signal $\beta_0$ and magnitude signal increase $\beta_1$, respectively\footnote{While not previously mentioned, now is a good time to acknowledge that SHAKER should not be used for quantitative MRI, as the final values have been through at least one layer of scaling and may not bear any quantitative \textit{meaning}.}. In addition to specifying \textit{SNR} and \textit{CNR}, users may choose to specify some amount of task-related-phase-change (TRPC) in degrees. It has been shown that there is biological information contained in the phase of an image, and as such it may be desirable to consider it in statistical models \citep{rowe-phase-2005}. In SHAKER, magnitude activation is determined through the \textit{CNR}, and an additional phase angle in the activated areas can be specified. Once all MRI parameters and fMRI options are set, users may select the \textit{Generate Time Series} button found in the \textit{Task Design} tab to initialize the simulation.

\subsection{Time series data analysis} \label{subsec:timeseries}
\begin{figure}[b!]
\centering
\includegraphics[width=0.5\linewidth]{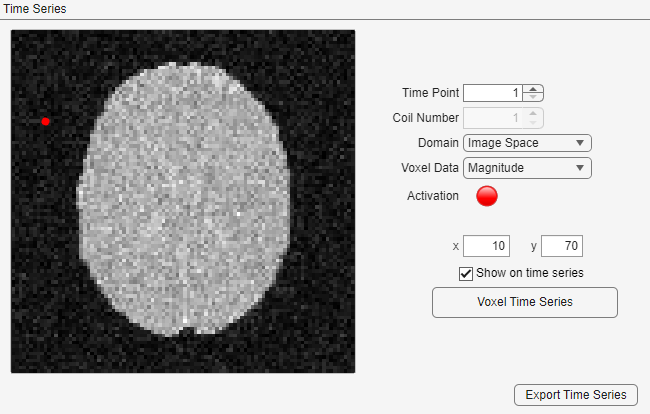}
\caption{\label{fig:ts-example} \textit{Time Series}: the top right pane of SHAKER. Here the simulated time series can be observed and checked for expected results.}
\end{figure}

The top right panel of SHAKER as in Figure \ref{fig:ts-example}, labeled as \textit{Time Series}, allows for visualization and examination of the simulated time series data. Users may look through individual images in the time series, in either $k$-space or image-space (if reconstructed), viewing the real, imaginary, magnitude, or phase parts of an image. There is also the option to monitor the time series of individual voxels which can be helpful to determine regions of activation and activation structure (magnitude/phase).

\begin{figure}[h!]
\centering
\includegraphics[width=0.5\linewidth]{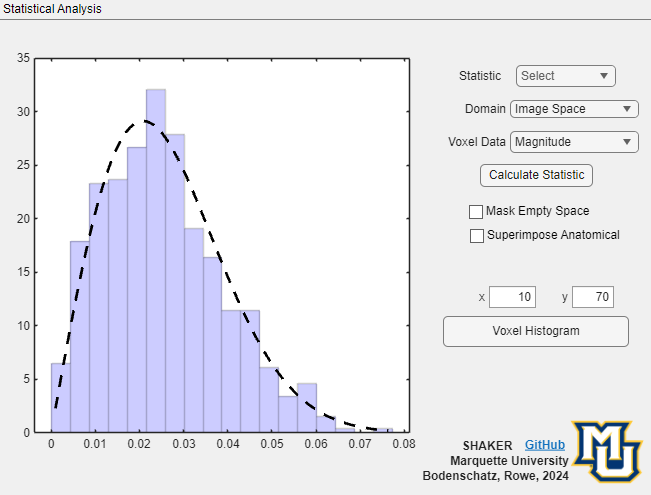}
\caption{\label{fig:stats-example} \textit{Statistical Analysis}: the bottom right pane of SHAKER. This is where users may analyze statistical properties from the data, such as: $t$~statistic for activation, $SNR$ maps, voxel histograms, etc.}
\end{figure}

Below this, in the bottom right panel of SHAKER as in Figure \ref{fig:stats-example}, labeled \textit{Statistical Analysis}, is where statistical maps and models of the simulated time series may be assessed. True to the MRI machine, SHAKER supports analysis of complex-valued data. Users may closely examine the real and imaginary part of an image, or magnitude and phase part of image. This can be done in both $k$-space and reconstructed image-space. SHAKER comes preloaded with two simple statistical measures: a voxel-wise $t$~test for activation detection and an \textit{SNR} calculation to evaluate image fidelity \citep{t-test-activation}. The $t$~test compares the task-active images in a simulated time series to the mean rest image in order to determine some change in magnitude or phase over some ROI. This is what is expected with the BOLD signal due to the increased $T_2^*$ effect. The \textit{SNR} calculation estimates the \textit{SNR} of each voxel throughout the time series. This helps determine the quality of $k$-space trajectory and reconstruction method by highlighting any regions of leakage or other artifacts. Statistical maps can be superimposed onto an anatomical image of the excited slice for better viewing of the activated regions or other ROIs. Additionally, there is the option to look at the histogram of any voxel's magnitude/phase/real/imaginary component throughout the time series with theoretical probability density functions (PDF) overlain. This may be used to confirm expected distributions of voxel's time series as described in Section \ref{subsec:fmri}. As described in the Appendix, SHAKER supports the use of custom statistical methods and models to analyze the simulated data. At present it is recommended that for advanced models the data be exported and examined in a more controlled environment.

\section{Example fMRI experiment} \label{sec:example}
This section will carry out an example simulated fMRI experiment, detailing all steps taken in SHAKER. Suppose that a right-hand finger tapping experiment was performed on a subject. The 3T machine is set to scan an axial slice in the center of the brain. The data matrix size is $96\times96$, with $TE = 60.4 ms$, $EESP = 0.832 ms$, $TR = 1s$, and no acceleration factor ($n_a=1$). The experimental timing starts with an initial 16s of rest followed by 19 epochs of 16s of task alternating with 16s of rest for a total of $n_{IMG}=624$ images.

Firstly, the phantom size and slice orientation and number would be set. This is done in the top left pane of SHAKER by setting the options \texttt{Phantom Size:\;96}, \texttt{View:\;Axial}, and \texttt{Slice:\;48}. Following this, the MRI parameters should be set. Without further knowledge of the experiment, it may be safe to assume that a GRE pulse sequence is used and measured along the standard Cartesian trajectory. So the options \texttt{Signal Equation:} \texttt{GradientEcho\_SigEq.m} and \texttt{Trajectory:\;Cartesian\_kspace.m} should be selected. Following this, the acceleration factor, field strength, \textit{TE}, \textit{TR}, and \textit{EESP} can be input directly from the experimental setup data. It can be assumed that the flip angle is $\alpha=90^\circ$ (this is not always the case experimentally, but, unless other information is known, is a reasonable assumption). For simplicity it can also be assumed that the machine is using a single, uniform coil. Since the $k$-space trajectory is the standard Cartesian path, images can be reconstructed by setting \texttt{Reconstruction:\;CartesianIFFT.m}. To better represent the machine, the box for $B_0$ inhomogeneity may be checked to include the $\Delta B$ effect into the simulation.

In the \textit{Task Design} tab next to \textit{MRI Parameters}, the options for the fMRI experimental design can be set. The HRF can be set to block and the four values that follow: initial rest images, epochs, and task/rest images per epoch, can be filled in directly from the experimental setup. The \textit{Plot Design} button can be pressed to visualize and ensure the experimental timing is setup correctly. To be consistent with empirical data it is recommended that the \textit{SNR} is set somewhere in the range of [1, 10] and the \textit{CNR} is set somewhere between [0.1, 2]. For this example, the two are set to be \texttt{SNR:\;5} and \texttt{CNR:\;0.5}. There will be no phase activation added in this example, so \texttt{Phase:\;0}. Once all inputs are confirmed to be correct, the time series is simulated by pressing the \textit{Generate Time Series Data} button.

\begin{figure}[t!]
\centering
\includegraphics[width=0.75\linewidth]{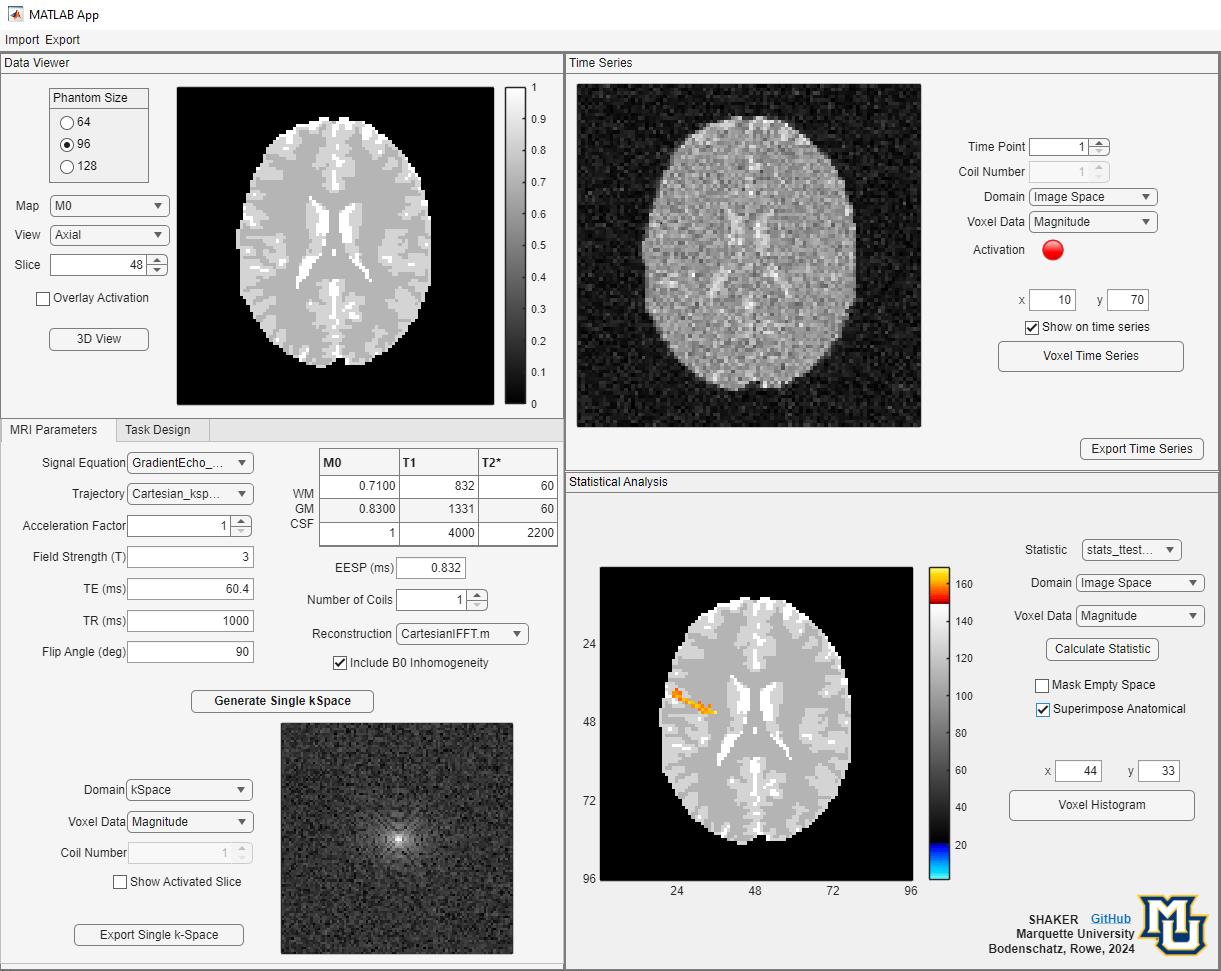}
\caption{\label{fig:example} Screenshot of the input to SHAKER for the example simulation described in Section \ref{sec:example}. Not pictured: the fMRI time series and \textit{SNR}/\textit{CNR} options set in the second tab of the bottom left panel.}
\end{figure}

Once the simulation is complete, a summary of simulation will be displayed and the \textit{Time Series} panel will be populated with data. From here, the images can be observed and sorted through to check for any errors in simulation. A time series of a voxel in an active region may be generated to observe any noticeable patterns. The data may then be analyzed using custom or built-in tools. This can be done for task detection, noise analysis, etc. The results for this example are shown in Figure \ref{fig:example}. The summary of simulation for this example reads:
\begin{quote}
    \textit{The following fMRI time series data was simulated on 11-Nov-2024 at 17:07:23. The simulated time series is of slice 48 from a size 96 phantom in the Axial plane. The MRI parameters were set to be the following: Acceleration Factor = 1, Field Strength = 3T, TE = 60.4ms, TR = 1000ms, Flip Angle = 90deg, EESP = 0.832ms, and Number of Coils = 1. The data was simulated with the GradientEcho\_SigEq.m signal equation using the Cartesian\_kspace.m k-space trajectory. The experimental design involved an initial 16 rest images followed by 19 epochs, each consisting of 16 task images followed by 16 rest images for a total of 624 images. The SNR was set to 5 and the CNR was set to 0.5. There were 0 degrees of phase added to the activation. Images were reconstructed using the CartesianIFFT.m algorithm.}
\end{quote}


\section{Discussion and future work} \label{sec:conclusion}
SHAKER is a one-stop shop for fMRI simulation. The GUI-based approach to the simulator allows for specially simulated data to get quickly into the hands of researchers as compared to long wait times for experimental data. The physics-based approach entrusts that the simulated data is representative of what a proper MRI machine might output. The most recent version of SHAKER can be found on GitHub at the following URL: \url{https://github.com/bodensjc/SHAKER}. Due to the public availability of SHAKER, it can remain in a constant state of development as users contribute ideas and needs for the simulation tool. All thoughts and considerations for future development are asked to be sent to the corresponding author.

Future consideration for this work involve the addition of more MRI features and parameters as well as additional fMRI experimental design components. Control of Field-of-View (FOV) and bandwidth could be helpful for generating zoomed images. More control of the noise generation through temporal variation as well as spatial. The inclusion of intra- and inter-imaging motion for fMRI experiments would help test registration and motion correction algorithms. Standard reconstruction methods such as SENSE and GRAPPA can be implemented to facilitate accelerated parallel imaging \citep{sense,grappa}. Proprietary non-uniform inverse Fourier transforms may also be implemented to facilitate the reconstruction of non-Cartesian based $k$-space trajectories. Simultaneous multi-slice (SMS) techniques such as CAIPIRINHA \citep{caipirinha} could be supported in the future as well as other full- or partial-volume imaging methods. Additional support for the processing of archival data will be added, including techniques such as: Nyquist ghost correction, zero-filling, apodization, motion correction, static  $B_0$  correction, etc.

\section{Acknowledgments}
The authors would like to thank the generosity of the Wehr Foundation and the Northwestern Mutual Data Science Institute for partially funding this work. The authors report there are no competing interests to declare.





 \bibliographystyle{elsarticle-harv} 
 \bibliography{fmri-refs}

\begin{thebibliography}{30}
\expandafter\ifx\csname natexlab\endcsname\relax\def\natexlab#1{#1}\fi
\providecommand{\url}[1]{\texttt{#1}}
\providecommand{\href}[2]{#2}
\providecommand{\path}[1]{#1}
\providecommand{\DOIprefix}{doi:}
\providecommand{\ArXivprefix}{arXiv:}
\providecommand{\URLprefix}{URL: }
\providecommand{\Pubmedprefix}{pmid:}
\providecommand{\doi}[1]{\href{http://dx.doi.org/#1}{\path{#1}}}
\providecommand{\Pubmed}[1]{\href{pmid:#1}{\path{#1}}}
\providecommand{\bibinfo}[2]{#2}
\ifx\xfnm\relax \def\xfnm[#1]{\unskip,\space#1}\fi
\bibitem[{Adrian et~al.(2013)Adrian, Maitra and
  Rowe}]{adrian-maitra-rowe-stat13}
\bibinfo{author}{Adrian, D.W.}, \bibinfo{author}{Maitra, R.},
  \bibinfo{author}{Rowe, D.B.}, \bibinfo{year}{2013}.
\newblock \bibinfo{title}{Ricean over gaussian modelling in magnitude {fMRI}
  {Analysis-Added} complexity with negligible practical benefits}.
\newblock \bibinfo{journal}{Stat} \bibinfo{volume}{2},
  \bibinfo{pages}{303--316}.
\bibitem[{Ardekani and Kanno(1998)}]{t-test-activation}
\bibinfo{author}{Ardekani, B.A.}, \bibinfo{author}{Kanno, I.},
  \bibinfo{year}{1998}.
\newblock \bibinfo{title}{Statistical methods for detecting activated regions
  in functional mri of the brain}.
\newblock \bibinfo{journal}{Magnetic Resonance Imaging} \bibinfo{volume}{16},
  \bibinfo{pages}{1217--1225}.
\newblock \URLprefix
  \url{https://www.sciencedirect.com/science/article/pii/S0730725X98001258},
  \DOIprefix\doi{https://doi.org/10.1016/S0730-725X(98)00125-8}.
\bibitem[{Bernstein et~al.(2004)Bernstein, King and
  Zhou}]{handbook-mri-pulse-sequences}
\bibinfo{author}{Bernstein, M.}, \bibinfo{author}{King, K.},
  \bibinfo{author}{Zhou, X.}, \bibinfo{year}{2004}.
\newblock \bibinfo{title}{Handbook of MRI Pulse Sequences}.
\newblock \bibinfo{publisher}{Elsevier Inc.}
\newblock \DOIprefix\doi{10.1016/B978-0-12-092861-3.X5000-6}.
  \bibinfo{note}{publisher Copyright: {\textcopyright} 2004 Elsevier Inc. All
  rights reserved.}
\bibitem[{Block et~al.(2014)Block, Chandarana, Milla, Bruno, Mulholland,
  Fatterpekar, Hagiwara, Grimm, Geppert, Kiefer and Sodickson}]{spiral}
\bibinfo{author}{Block, K.}, \bibinfo{author}{Chandarana, H.},
  \bibinfo{author}{Milla, S.}, \bibinfo{author}{Bruno, M.},
  \bibinfo{author}{Mulholland, T.}, \bibinfo{author}{Fatterpekar, G.},
  \bibinfo{author}{Hagiwara, M.}, \bibinfo{author}{Grimm, R.},
  \bibinfo{author}{Geppert, C.}, \bibinfo{author}{Kiefer, B.},
  \bibinfo{author}{Sodickson, D.}, \bibinfo{year}{2014}.
\newblock \bibinfo{title}{Towards routine clinical use of radial stack-of-stars
  3d gradient-echo sequences for reducing motion sensitivity}.
\newblock \bibinfo{journal}{Journal of the Korean Society of Magnetic Resonance
  in Medicine} \bibinfo{volume}{18}, \bibinfo{pages}{87}.
\newblock \DOIprefix\doi{10.13104/jksmrm.2014.18.2.87}.
\bibitem[{Breuer et~al.(2005)Breuer, Blaimer, Heidemann, Mueller, Griswold and
  Jakob}]{caipirinha}
\bibinfo{author}{Breuer, F.A.}, \bibinfo{author}{Blaimer, M.},
  \bibinfo{author}{Heidemann, R.M.}, \bibinfo{author}{Mueller, M.F.},
  \bibinfo{author}{Griswold, M.A.}, \bibinfo{author}{Jakob, P.M.},
  \bibinfo{year}{2005}.
\newblock \bibinfo{title}{Controlled aliasing in parallel imaging results in
  higher acceleration (caipirinha) for multi-slice imaging}.
\newblock \bibinfo{journal}{Magnetic Resonance in Medicine}
  \bibinfo{volume}{53}, \bibinfo{pages}{684--691}.
\newblock \DOIprefix\doi{https://doi.org/10.1002/mrm.20401}.
\bibitem[{Chen and Glover(2015)}]{Chen2015}
\bibinfo{author}{Chen, J.E.}, \bibinfo{author}{Glover, G.H.},
  \bibinfo{year}{2015}.
\newblock \bibinfo{title}{Functional magnetic resonance imaging methods}.
\newblock \bibinfo{journal}{Neuropsychology review} \bibinfo{volume}{25},
  \bibinfo{pages}{289--313}.
\newblock \DOIprefix\doi{https://doi.org/10.1007/s11065-015-9294-9}.
\bibitem[{Comby et~al.(2024)Comby, Vignaud and Ciuciu}]{snake-fmri}
\bibinfo{author}{Comby, P.A.}, \bibinfo{author}{Vignaud, A.},
  \bibinfo{author}{Ciuciu, P.}, \bibinfo{year}{2024}.
\newblock \bibinfo{title}{Snake-fmri: A modular fmri data simulator from the
  space-time domain to k-space and back}.
\newblock \URLprefix \url{https://arxiv.org/abs/2404.08282},
  \href{http://arxiv.org/abs/2404.08282}{{\tt arXiv:2404.08282}}.
\bibitem[{Elster et~al.(2001)Elster, Burdette and Field}]{questions}
\bibinfo{author}{Elster, A.D.}, \bibinfo{author}{Burdette, J.H.},
  \bibinfo{author}{Field, A.}, \bibinfo{year}{2001}.
\newblock \bibinfo{title}{Questions {\&} answers in magnetic resonance
  imaging}.
\newblock \bibinfo{edition}{Second edition.} ed., \bibinfo{publisher}{Mosby},
  \bibinfo{address}{St. Louis}.
\bibitem[{Griswold et~al.(2002)Griswold, Jakob, Heidemann, Nittka, Jellus,
  Wang, Kiefer and Haase}]{grappa}
\bibinfo{author}{Griswold, M.A.}, \bibinfo{author}{Jakob, P.M.},
  \bibinfo{author}{Heidemann, R.M.}, \bibinfo{author}{Nittka, M.},
  \bibinfo{author}{Jellus, V.}, \bibinfo{author}{Wang, J.},
  \bibinfo{author}{Kiefer, B.}, \bibinfo{author}{Haase, A.},
  \bibinfo{year}{2002}.
\newblock \bibinfo{title}{Generalized autocalibrating partially parallel
  acquisitions (grappa)}.
\newblock \bibinfo{journal}{Magnetic Resonance in Medicine}
  \bibinfo{volume}{47}, \bibinfo{pages}{1202--1210}.
\newblock \URLprefix
  \url{https://onlinelibrary.wiley.com/doi/abs/10.1002/mrm.10171},
  \DOIprefix\doi{https://doi.org/10.1002/mrm.10171},
  \href{http://arxiv.org/abs/https://onlinelibrary.wiley.com/doi/pdf/10.1002/mrm.10171}{{\tt
  arXiv:https://onlinelibrary.wiley.com/doi/pdf/10.1002/mrm.10171}}.
\bibitem[{Gudbjartsson and Patz(1995)}]{rician-noise}
\bibinfo{author}{Gudbjartsson, H.}, \bibinfo{author}{Patz, S.},
  \bibinfo{year}{1995}.
\newblock \bibinfo{title}{The rician distribution of noisy mri data}.
\newblock \bibinfo{journal}{Magnetic Resonance in Medicine}
  \bibinfo{volume}{34}, \bibinfo{pages}{910--914}.
\newblock \URLprefix
  \url{https://onlinelibrary.wiley.com/doi/abs/10.1002/mrm.1910340618},
  \DOIprefix\doi{https://doi.org/10.1002/mrm.1910340618},
  \href{http://arxiv.org/abs/https://onlinelibrary.wiley.com/doi/pdf/10.1002/mrm.1910340618}{{\tt
  arXiv:https://onlinelibrary.wiley.com/doi/pdf/10.1002/mrm.1910340618}}.
\bibitem[{Haase et~al.(1986)Haase, Frahm, Matthaei, Hanicke and
  Merboldt}]{FLASH}
\bibinfo{author}{Haase, A.}, \bibinfo{author}{Frahm, J.},
  \bibinfo{author}{Matthaei, D.}, \bibinfo{author}{Hanicke, W.},
  \bibinfo{author}{Merboldt, K.D.}, \bibinfo{year}{1986}.
\newblock \bibinfo{title}{Flash imaging. rapid nmr imaging using low flip-angle
  pulses}.
\newblock \bibinfo{journal}{Journal of Magnetic Resonance (1969)}
  \bibinfo{volume}{67}, \bibinfo{pages}{258--266}.
\newblock \URLprefix
  \url{https://www.sciencedirect.com/science/article/pii/0022236486904336},
  \DOIprefix\doi{https://doi.org/10.1016/0022-2364(86)90433-6}.
\bibitem[{Henkelman(1985)}]{henkelman}
\bibinfo{author}{Henkelman, R.M.}, \bibinfo{year}{1985}.
\newblock \bibinfo{title}{Measurement of signal intensities in the presence of
  noise in mr images}.
\newblock \bibinfo{journal}{Medical Physics} \bibinfo{volume}{12},
  \bibinfo{pages}{232--233}.
\newblock \URLprefix
  \url{https://aapm.onlinelibrary.wiley.com/doi/abs/10.1118/1.595711},
  \DOIprefix\doi{https://doi.org/10.1118/1.595711}.
\bibitem[{Karaman(2014)}]{Karaman_2014_phd}
\bibinfo{author}{Karaman, M.}, \bibinfo{year}{2014}.
\newblock \bibinfo{title}{Improving fMRI Analysis and MR Reconstruction With
  the Incorporation of MR Relaxivities and Correlation Effect Examination}.
\newblock Ph.D. thesis. Marquette University.
\bibitem[{Karaman et~al.(2015)Karaman, Bruce and Rowe}]{KaramanBruceRoweMRI15}
\bibinfo{author}{Karaman, M.}, \bibinfo{author}{Bruce, I.},
  \bibinfo{author}{Rowe, D.}, \bibinfo{year}{2015}.
\newblock \bibinfo{title}{Incorporating relaxivities to more accurately
  reconstruct {MR} images}.
\newblock \bibinfo{journal}{Magnetic Resonance Imaging} \bibinfo{volume}{33},
  \bibinfo{pages}{374--384}.
\newblock \URLprefix
  \url{https://www.sciencedirect.com/science/article/pii/S0730725X15000041},
  \DOIprefix\doi{https://doi.org/10.1016/j.mri.2015.01.003}.
\bibitem[{Kida et~al.(2000)Kida, Kennan, Rothman, Behar and
  Hyder}]{kennan-spin-echo}
\bibinfo{author}{Kida, I.}, \bibinfo{author}{Kennan, R.P.},
  \bibinfo{author}{Rothman, D.L.}, \bibinfo{author}{Behar, K.L.},
  \bibinfo{author}{Hyder, F.}, \bibinfo{year}{2000}.
\newblock \bibinfo{title}{High-resolution cmro2 mapping in rat cortex: A
  multiparametric approach to calibration of bold image contrast at 7 tesla}.
\newblock \bibinfo{journal}{Journal of Cerebral Blood Flow \& Metabolism}
  \bibinfo{volume}{20}, \bibinfo{pages}{847--860}.
\newblock \DOIprefix\doi{10.1097/00004647-200005000-00012}.
  \bibinfo{note}{pMID: 10826536}.
\bibitem[{Koay and Basser(2006)}]{Koay2006}
\bibinfo{author}{Koay, C.G.}, \bibinfo{author}{Basser, P.J.},
  \bibinfo{year}{2006}.
\newblock \bibinfo{title}{Analytically exact correction scheme for signal
  extraction from noisy magnitude mr signals.}
\newblock \bibinfo{journal}{Journal of magnetic resonance} \bibinfo{volume}{179
  2}, \bibinfo{pages}{317--22}.
\newblock \URLprefix \url{https://api.semanticscholar.org/CorpusID:17979879}.
\bibitem[{Kumar et~al.(1975)Kumar, Welti and Ernst}]{KUMAR1975}
\bibinfo{author}{Kumar, A.}, \bibinfo{author}{Welti, D.},
  \bibinfo{author}{Ernst, R.R.}, \bibinfo{year}{1975}.
\newblock \bibinfo{title}{Nmr fourier zeugmatography}.
\newblock \bibinfo{journal}{Journal of Magnetic Resonance (1969)}
  \bibinfo{volume}{18}, \bibinfo{pages}{69--83}.
\newblock \URLprefix
  \url{https://www.sciencedirect.com/science/article/pii/0022236475902243},
  \DOIprefix\doi{https://doi.org/10.1016/0022-2364(75)90224-3}.
\bibitem[{Larmor(1897)}]{larmor}
\bibinfo{author}{Larmor, J.}, \bibinfo{year}{1897}.
\newblock \bibinfo{title}{Ix. a dynamical theory of the electric and
  luminiferous medium.— part iii. relations with material media}.
\newblock \bibinfo{journal}{Philosophical Transactions of the Royal Society of
  London. Series A, Containing Papers of a Mathematical or Physical Character}
  \bibinfo{volume}{190}, \bibinfo{pages}{205--300}.
\newblock \URLprefix
  \url{https://royalsocietypublishing.org/doi/abs/10.1098/rsta.1897.0020},
  \DOIprefix\doi{10.1098/rsta.1897.0020},
  \href{http://arxiv.org/abs/https://royalsocietypublishing.org/doi/pdf/10.1098/rsta.1897.0020}{{\tt
  arXiv:https://royalsocietypublishing.org/doi/pdf/10.1098/rsta.1897.0020}}.
\bibitem[{Lindquist(2008)}]{noise}
\bibinfo{author}{Lindquist, M.A.}, \bibinfo{year}{2008}.
\newblock \bibinfo{title}{The statistical analysis of fmri data}.
\newblock \bibinfo{journal}{Statistical Science} \bibinfo{volume}{23},
  \bibinfo{pages}{439 -- 464}.
\newblock \URLprefix \url{https://doi.org/10.1214/09-STS282},
  \DOIprefix\doi{10.1214/09-STS282}.
\bibitem[{Nencka and Rowe(2005)}]{nencka-rowe-SE-2005}
\bibinfo{author}{Nencka, A.S.}, \bibinfo{author}{Rowe, D.B.},
  \bibinfo{year}{2005}.
\newblock \bibinfo{title}{Complex constant phase activation model removes
  venous bold contribution in fmri}.
\newblock \bibinfo{journal}{The International Society for Magnetic Resonance in
  Medicine} .
\bibitem[{Ogawa et~al.(1990)Ogawa, Lee, Kay and Tank}]{BOLD}
\bibinfo{author}{Ogawa, S.}, \bibinfo{author}{Lee, T.M.}, \bibinfo{author}{Kay,
  A.R.}, \bibinfo{author}{Tank, D.W.}, \bibinfo{year}{1990}.
\newblock \bibinfo{title}{Brain magnetic resonance imaging with contrast
  dependent on blood oxygenation}.
\newblock \bibinfo{journal}{Proc. Natl. Acad. Sci. U. S. A.}
  \bibinfo{volume}{87}, \bibinfo{pages}{9868--9872}.
\bibitem[{Pipe(1999)}]{PROPELLER}
\bibinfo{author}{Pipe, J.G.}, \bibinfo{year}{1999}.
\newblock \bibinfo{title}{Motion correction with propeller mri: Application to
  head motion and free-breathing cardiac imaging}.
\newblock \bibinfo{journal}{Magnetic Resonance in Medicine}
  \bibinfo{volume}{42}, \bibinfo{pages}{963--969}.
\newblock
  \DOIprefix\doi{https://doi.org/10.1002/(SICI)1522-2594(199911)42:5<963::AID-MRM17>3.0.CO;2-L}.
\bibitem[{Pruessmann et~al.(1999)Pruessmann, Weiger, Scheidegger and
  Boesiger}]{sense}
\bibinfo{author}{Pruessmann, K.P.}, \bibinfo{author}{Weiger, M.},
  \bibinfo{author}{Scheidegger, M.B.}, \bibinfo{author}{Boesiger, P.},
  \bibinfo{year}{1999}.
\newblock \bibinfo{title}{Sense: Sensitivity encoding for fast mri}.
\newblock \bibinfo{journal}{Magnetic Resonance in Medicine}
  \bibinfo{volume}{42}, \bibinfo{pages}{952--962}.
\newblock
  \DOIprefix\doi{https://doi.org/10.1002/(SICI)1522-2594(199911)42:5<952::AID-MRM16>3.0.CO;2-S}.
\bibitem[{Rayleigh(1880)}]{rayleigh-distribution}
\bibinfo{author}{Rayleigh, L.}, \bibinfo{year}{1880}.
\newblock \bibinfo{title}{Xii. on the resultant of a large number of vibrations
  of the same pitch and of arbitrary phase}.
\newblock \bibinfo{journal}{The London, Edinburgh, and Dublin Philosophical
  Magazine and Journal of Science} \bibinfo{volume}{10},
  \bibinfo{pages}{73--78}.
\newblock \URLprefix \url{https://doi.org/10.1080/14786448008626893},
  \DOIprefix\doi{10.1080/14786448008626893},
  \href{http://arxiv.org/abs/https://doi.org/10.1080/14786448008626893}{{\tt
  arXiv:https://doi.org/10.1080/14786448008626893}}.
\bibitem[{Rice(1944)}]{rice-distribution}
\bibinfo{author}{Rice, S.O.}, \bibinfo{year}{1944}.
\newblock \bibinfo{title}{Mathematical analysis of random noise}.
\newblock \bibinfo{journal}{Bell System Technical Journal}
  \bibinfo{volume}{23}, \bibinfo{pages}{282--332}.
\newblock \URLprefix
  \url{https://onlinelibrary.wiley.com/doi/abs/10.1002/j.1538-7305.1944.tb00874.x},
  \DOIprefix\doi{https://doi.org/10.1002/j.1538-7305.1944.tb00874.x},
  \href{http://arxiv.org/abs/https://onlinelibrary.wiley.com/doi/pdf/10.1002/j.1538-7305.1944.tb00874.x}{{\tt
  arXiv:https://onlinelibrary.wiley.com/doi/pdf/10.1002/j.1538-7305.1944.tb00874.x}}.
\bibitem[{Rowe(2005)}]{rowe-phase-2005}
\bibinfo{author}{Rowe, D.B.}, \bibinfo{year}{2005}.
\newblock \bibinfo{title}{Modeling both magnitude and phase of complex {fMRI}
  data}.
\newblock \bibinfo{journal}{NeuroImage} \bibinfo{volume}{25},
  \bibinfo{pages}{1310--24}.
\newblock \DOIprefix\doi{10.1016/j.neuroimage.2005.01.034}.
\bibitem[{Rowe(2016)}]{handbook-chapter-rowe}
\bibinfo{author}{Rowe, D.B.}, \bibinfo{year}{2016}.
\newblock \bibinfo{title}{Image Reconstruction in Functional MRI}.
\newblock \bibinfo{edition}{First edition.} ed., \bibinfo{publisher}{CRC
  Press}.
\newblock \URLprefix \url{https://doi.org/10.1201/9781315373652},
  \DOIprefix\doi{10.1201/9781315373652}.
\bibitem[{Rowe(2023)}]{rowe-jsm-2023}
\bibinfo{author}{Rowe, D.B.}, \bibinfo{year}{2023}.
\newblock \bibinfo{title}{Statistics of intrinsic fmri data}.
\newblock \URLprefix \url{http://dx.doi.org/10.5281/zenodo.10002334},
  \DOIprefix\doi{10.5281/zenodo.10002334}.
\bibitem[{{The MathWorks Inc.}(2024)}]{MATLAB}
\bibinfo{author}{{The MathWorks Inc.}}, \bibinfo{year}{2024}.
\newblock \bibinfo{title}{Matlab version: 24.1.0 (r2024a)}.
\newblock \URLprefix \url{https://www.mathworks.com}.
\bibitem[{Welvaert et~al.(2011)Welvaert, Durnez, Moerkerke, Berdoolaege and
  Rosseel}]{neurosim}
\bibinfo{author}{Welvaert, M.}, \bibinfo{author}{Durnez, J.},
  \bibinfo{author}{Moerkerke, B.}, \bibinfo{author}{Berdoolaege, G.},
  \bibinfo{author}{Rosseel, Y.}, \bibinfo{year}{2011}.
\newblock \bibinfo{title}{neurosim: An {R} package for generating fmri data}.
\newblock \bibinfo{journal}{Journal of Statistical Software}
  \bibinfo{volume}{44}, \bibinfo{pages}{1–18}.
\newblock \URLprefix
  \url{https://www.jstatsoft.org/index.php/jss/article/view/v044i10},
  \DOIprefix\doi{10.18637/jss.v044.i10}.

\end{thebibliography}








\end{document}